\documentclass[aps,prd,showpacs,preprintnumbers]{revtex4}
\usepackage{graphicx}
\usepackage{epsfig}
\usepackage{color}
\usepackage{amsmath}
\usepackage{epsf,amssymb,amsfonts,amsmath}
\newcommand{\bs}{\boldsymbol}
\newcommand{\ext}{{\mathrm{ext}}}
\newcommand{\eq}[1]{(\ref{#1})}
\newcommand{\be}{\begin{eqnarray}}
\newcommand{\ee}{\end{eqnarray}}

\newcommand{\ep}{\epsilon}
\newcommand{\bea}{\begin{eqnarray}}
\newcommand{\eea}{\end{eqnarray}}

\begin{document}

\title{Possible formation of high temperature superconductor at early stage of heavy-ion collisions}
\author{Hao Liu $^{a}$}
\author{Lang Yu$^{b}$}
\author{Maxim Chernodub$^{c}$}
\author{Mei Huang$^{a,d}$}
\affiliation{$^{a}$  Institute of High Energy Physics, Chinese Academy of Sciences, Beijing 100049, P.R. China}
\affiliation{$^{b}$ Center of Theoretical Physics and College of Physics, Jilin University, Changchun, 130012, P.R. China}
\affiliation{$^{c}$ CNRS, Laboratoire de Math\'ematiques et Physique Th\'eorique UMR 7350, Universit\'e de Tours, 37200 France }
\affiliation{$^{d}$ Theoretical Physics Center for Science Facilities, Chinese Academy of Sciences, Beijing 100049, P.R. China}

\begin{abstract}
We investigate the effect of the inverse magnetic catalysis (IMC) on the charged $\rho$ meson condensation at finite temperature in the framework of the Nambu--Jona-Lasinio model, where mesons are calculated to the leading order of $1/N_c$ expansion by summing up infinity quark-loops. IMC for chiral condensate has been considered in three different ways, i.e. fitting Lattice data, using the running coupling constant and  introducing the chiral chemical potential, respectively. It is observed that, with no IMC effect included, the critical magnetic field  $eB_c$ for charged $\rho$ condensation increases monotonically with the temperature. However,  including IMC substantially affects the polarized charged $\rho$ condensation around the critical temperature $T_c$ of chiral phase transition,  the critical magnetic field $eB_c$ for charged $\rho$ condensation decreases with the temperature firstly, reaches to a minimum value around $T_c$, then increases with the temperature. Our calculation indicates that the charged $\rho$ condensation can exist in the temperature region of $1-1.5 T_c$ with critical magnetic field $eB_c\sim 0.15-0.3 {\rm GeV}^2$, which suggests that high temperature superconductor might be created through non-central heavy ion collisions at LHC energies. We also show that a growing electric conductivity in early stage of non-central heavy-ion collisions substantially delays the decay of strong magnetic field, which is helpful for the formation of the high temperature superconductor.
\end{abstract}
\maketitle

\section{Introduction}

Quantum Chromodynamics (QCD) is widely believed to be the fundamental theory of the strong interactions. Investigations on its rich vacuum structure and how the QCD vacuum can be modified in an extreme environment are among the major theoretical challenges in modern physics. The extreme environment includes high temperature, finite baryonic chemical potential,  and recently more interests are attracted to strong magnetic fields \cite{Andersen:2014xxa,Miransky:2015ava,Huang:2015oca}.
In the surface of magnetars, magnetic fields can reach $10^{14-15} ~G$, which is a thousand times larger than that of an average pulsar, and in the inner core of magnetars the magnetic fields could reach as high as $10^{18}\sim 10^{20}~\textmd{G}$ \cite{Magnetic-NeutronStar}. Strong magnetic fields with strength of $10^{18}\sim 10^{20}  ~\textmd{G}$ [corresponding to $eB\sim (0.1-1.0~{\rm GeV})^2$], can be generated in the laboratory through non-central heavy ion collisions \cite{Skokov:2009qp,Deng:2012pc} at the Relativistic Heavy Ion Collider (RHIC) and the Large Hadron Collider (LHC). Hence, the heavy ion collisions experiment supplies us a good platform to investigate the QCD vacuum and matter under strong magnetic fields. Strong magnetic fields cause many interesting effects on properties of quark matter. For example, the chiral magnetic effect (CME) \cite{Kharzeev:2007tn,Kharzeev:2007jp,Fukushima:2008xe}, chiral vortical effect (CVE) \cite{Kharzeev:2010gr}, the magnetic catalysis (MC) \cite{Klevansky:1989vi,Klimenko:1990rh,Gusynin:1995nb} and inverse magnetic catalysis (IMC) \cite{Bali:2011qj, Bali:2012zg, Bali:2013esa}, and the formation vacuum superconductor \cite{Chernodub:2010qx,Chernodub:2011mc} etc.

Spontaneous breaking of chiral symmetry is one of the most important properties of strong interactions. It is well known that at zero chemical potential the chiral symmetry is broken in the low-temperature phase of QCD and it is restored at high temperature. The influence of the external magnetic field on chiral symmetry breaking and restoration has been a subject of intensive studies for several years. At low temperature, the background magnetic field is known to enhance the chiral symmetry breaking by increasing the value of the chiral condensate~\cite{Klevansky:1989vi,Klimenko:1990rh,Gusynin:1995nb}. This phenomenon, known as the magnetic catalysis, was confirmed in various effective low-energy models of QCD~\cite{Klevansky:1989vi, Klimenko:1990rh, Gusynin:1995nb, Shushpanov:1997sf, Agasian:1999sx, Alexandre:2000yf,
Agasian:2001hv, Cohen:2007bt, Gatto:2010qs, Gatto:2010pt, Mizher:2010zb, Kashiwa:2011js,
Avancini:2012ee, Andersen:2012dz, Scherer:2012nn} as well as in numerical simulations of lattice QCD~\cite{Buividovich:2008wf,
Braguta:2010ej, D'Elia:2010nq, D'Elia:2011zu,Ilgenfritz:2012fw}. Since at low temperature the background magnetic field and thermal fluctuations have opposite effects on the chiral symmetry by, respectively, increasing and decreasing the chiral condensate, then one can naturally assume that increasing background magnetic field should lead to increase of the critical temperature $T_c$ of the chiral symmetry restoration. Early analytical calculations in the most simplest effective QCD models~\cite{Agasian:2001hv, Cohen:2007bt, Gatto:2010qs, Gatto:2010pt, Mizher:2010zb, Kashiwa:2011js,
Avancini:2012ee, Andersen:2012dz, Scherer:2012nn} and lattice simulations with heavy quarks~\cite{Buividovich:2008wf, D'Elia:2011zu} have suggested that the Magnetic Catalysis should also be realized at high temperature as well. However, detailed QCD calculations with realistic quark masses have revealed that the chiral transition temperature $T_c$ is, unexpectedly, a decreasing function of the background magnetic field~\cite{Bali:2011qj, Bali:2012zg, Bali:2013esa}. This phenomenon is now known as the inverse magnetic catalysis (IMC).

Theoretical understanding of the mechanism of the IMC has sparked significant activity and further controversies. At the moment there is no unified opinion on the physical mechanism that underlines the IMC. For example, in Ref.~\cite{Fukushima:2012kc} the IMC was suggested to be caused by a magnetic inhibition effect related to the anisotropy of neutral meson propagation. In Ref.~\cite{Kamikado:2013pya} the magnetic inhibition was however shown to be ineffective to explain the IMC and  it was recently shown in \cite{Mao:2016fha} that the IMC might be induced by neutral pion fluctuation if Pauli-villas regularization is used. In Ref.~\cite{Kojo:2012js} the IMC was proposed to be related to particularities of the infrared contributions to the quark mass gap, while in Ref.~\cite{Bruckmann:2013oba} the IMC is explained by different effects of magnetic field on the low quark modes -- that are responsible for the chiral symmetry breaking -- coming from valence and sea quarks. Alternatively, the functional renormalization group study of Ref.~\cite{Mueller:2015fka} have shown  that the physics underlying the IMC at high temperature cannot be captured by the standard Nambu--Jona-Lasinio (NJL) model, that is one of the most successful models of the chiral sector of low energy QCD. The IMC may, however, be reproduced if one introduces a phenomenological dependence of the NJL coupling on the magnetic field~\cite{Ferreira:2014kpa}.

Yet another interesting opportunity to explain the IMC comes from the fact that the theory of strong interactions has a nontrivial topological structure due to  existence of certain gluon configurations, instantons~\cite{Belavin:1975fg,tHooft:1976up,tHooft:1976fv} and their thermal cousins, sphalerons~\cite{Manton:1983nd,Klinkhamer:1984di,Kuzmin:1985mm,Arnold:1987mh,Khlebnikov:1988sr,Arnold:1987zg}, which describe transitions of QCD vacuum between different topological sectors. The chirality imbalance induced by the sphaleron transition~\cite{Chao:2013qpa} or instanton-anti-instanton molecule pairing~\cite{Yu:2014sla} can be invoked to explain the IMC around at high temperature. In continuation of these studies, the effect of the chiral chemical potential $\mu_5$ -- which characterizes (local) imbalance between fermions with different chirality -- on the chiral phase transition in the NJL model with different regulation schemes was investigated in the Ref.~\cite{Yu:2015hym}.

Another interesting effect that may be realized in strong magnetic field is the electromagnetic superconductivity of the QCD vacuum~\cite{Chernodub:2010qx,Chernodub:2011mc}. The idea is based on the simple fact that the energy levels of a free pointlike charged particle in a static uniform external magnetic field $B$ are $\varepsilon_{n,s_z}^2(p_z) = p_z^2+(2 n - 2 \, \text{sign}(q) s_z + 1) |qB| + m^2$, where $q$ is the electric charge of the particle, $n \geqslant 0$ is the nonnegative integer that characterizes the Landau levels, $s_z$ is the projection of particle's spin on the magnetic field axis $z$, and $p_z$ is particle's momentum along the magnetic field. The ground-state mass $M_{\rho^\pm}(B) = \sqrt{m_{\rho^\pm}^2 - |eB|}$ of a charged $\rho^{\pm}$ meson with the unit spin $s = 1$ corresponds to the lowest energy level with quantum numbers $p_z = 0$, $n = 0$ and $s_z = {\mathrm{sign}} (q)$. As the magnetic field increases, the $\rho^{\pm}$ mass decreases to zero that implies the instability of the ground state towards the condensation of the charged $\rho$ mesons at the critical magnetic field $eB_c = m_{\rho^\pm}^2\approx 0.6$ GeV$^2$.

The idea of the magnetic-field-induced vacuum superconductor is still under debates as different approaches give controversial results. The NJL model approach~\cite{Chernodub:2011mc,Frasca:2013kka,Liu:2014uwa} and independent non-perturbative holographic AdS/QCD techniques~\cite{Callebaut:2011uc,Ammon:2011je} indicate that the mass of the $\rho$-meson excitation should indeed vanish at certain value of the magnetic field. On the other hand, the relativistic Hamiltonian technique~\cite{Andreichikov:2013zba} and the Dyson-Schwinger equations~\cite{Wangkunlun:2013} lead to the conclusion that the mass of charged $\rho$ meson decreases at small magnetic field and then it increases as magnetic field becomes larger while never dropping to zero. Another approach based on hidden local symmetry~\cite{Kawaguchi:2015gpt} at $O(p^2)$ and $O(p^4)$ orders favors the vanishing of the $\rho$-meson mass supporting the conclusion obtained in the $\rho$-meson electrodynamics~\cite{Chernodub:2010qx} while the $O(p^6)$ results are inconclusive. On the contrary, the existing numerical calculations in quenched lattice QCD without dynamical quarks~\cite{Hidaka:2012mz,Luschevskaya:2014mna} show that the mass of the $\rho^\pm$ does not vanish while the calculations in lattice QCD with light dynamical quarks have not been done so far.

In Ref.~\cite{Hidaka:2012mz} it was argued that the $\rho$-meson condensation is forbidden by the Vafa-Witten (VW) theorem because the VW theorem implies the absence of any massless Nambu-Goldstone (NG) bosons in QCD except for the pions (the latter are associated with the chiral symmetry breaking). According to Ref.~\cite{Hidaka:2012mz}, if the $\rho$-mesons are condensed in QCD then they would lead to appearance of an NG boson which is forbidden by the VW theorem. This argument was refuted in Ref.~\cite{Chernodub:2012zx} where it was demonstrated that the $\rho$-meson condensation in external magnetic field takes place not in QCD but in QCD$\times$QED and, consequently, the would-be NG bosons associated with the $\rho$-meson condensation are absorbed into the longitudinal component of the electromagnetic gauge field in a Higgs-like mechanism. Consequently, no NG bosons appear due to the $\rho$-meson condensation in agreement with the VW theorem. A similar conclusion about the consistency of the $\rho$ condensation with the VW theorem was put forward in Ref.~\cite{Li:2013aa}. Moreover, $\rho$ mass calculations  in quenched lattice QCD~\cite{Hidaka:2012mz,Luschevskaya:2014mna} are consistent with a crossover transition to the superconducting phase~\cite{Chernodub:2013uja}.

In  Ref.\cite{Liu:2014uwa}, by calculating the $\rho$ meson polarization function to the leading order of $1/N_c$ expansion in the NJL model, we have obtained the critical magnetic field $eB_c\approx 0.2 {\rm GeV}^2$, which is only 1/3 of the results from the point-particle calculation in \cite{Chernodub:2010qx}. Furthermore, in Ref.~\cite{Liu:2015pna}, it was observed that in the temperature range $0.2-0.5~ {\rm GeV}$, the critical magnetic field for charged $\rho$ condensation is in the range of $0.2-0.6~ {\rm GeV}^2$, which indicates that the high temperature electromagnetic superconductor could be created at LHC. However, in Ref.~\cite{Liu:2015pna}, we have used the MC for chiral condensate to produce the constituent quark mass. In this work, we will consider the effect of the IMC on the charged $\rho$ meson condensate. This paper is organized as follows:  In Sec. II, By taking the quark propagator in the Landau level representation, we give a general description of the two-flavor NJL model including the effective four-quark interaction in the vector channel, and derive the vector meson mass under magnetic fields at finite temperature and chemical potential. We introduce three different mechanisms for IMC in Sec.III and investigate how the IMC will affect the charged $\rho$ meson condensation. In Sec. IV, we show the effect of large electric conductivity of high temperature conductor on the decay of strong magnetic field. In Sec. V the discussion and conclusion is given.

\section{NJL model and $\rho$ meson construction}

\subsection{The SU(2) magnetized NJL model}

We investigate the charged $\rho$ meson condensation in the magnetic field background at nonzero temperature in the framework of a two-flavor NJL model. The Lagrangian density of the NJL model takes the following form \cite{Nambu:1961tp,Nambu:1961fr,Klimt:1989pm,Vogl:1991qt,Klevansky:1992qe,Hatsuda:1994pi}:
\begin{eqnarray}
  {\cal{L}}&=&\bar{\psi}(i\not{\!\!D}-\hat{m})\psi+G_{S}\left[(\bar{\psi}\psi)^2
   +(\bar{\psi}i\gamma^5\vec{\tau}\psi)^2\right] \nonumber \\
 & & -G_{V}\left[(\bar{\psi}\gamma^{\mu}\mathbf{\tau}^a\psi)^2+
  (\bar{\psi}\gamma^{\mu}\gamma^5\mathbf{\tau}^a\psi)^2 \right] -\frac{1}{4}F_{\mu\nu}F^{\mu\nu}\,,
\label{eq:L:basic}
\end{eqnarray}
where $\psi = (u,d)^T$ is the doublet of the two light quark flavors $u$ and $d$ with the masses given by the diagonal current mass matrix $\hat{m}={\mathrm{diag}}(m_u, m_d)$ and $\tau^a=({\mathbb{I}}, \vec{\tau})$ with $\vec{\tau}=(\tau^1, \tau^2, \tau^3)$ are the isospin Pauli matrices. $G_S$ and $G_V$ are the coupling constants with respect to the scalar (pseudoscalar) and the vector (axial-vector) channels, respectively. The covariant derivative  $D_\mu=\partial_\mu - i q_f A_{\mu}^{\ext}$ couples the quarks to an external magnetic ${\bs B}=(0,0,B)$ along the positive $z$ direction, via a background field $A^\ext_\mu = -\delta_{\mu1}Bx_2$. The electric charges of the $(u,d)$ quark fields are $q_f=(-e/3, 2e/3)$ and the field strength tensor is $F_{\mu\nu}=\partial_{[\mu}A_{\nu]}^{\ext}$.

The four-point interactions in Eq.~(\ref{eq:L:basic}) can be semi-bosonized with respect to a Gaussian transformations and the NJL Lagrangian~\eq{eq:L:basic} can be rewritten in the following form:
\begin{eqnarray}\label{NE2b}
{\cal{L}}_{sb}&=&\bar{\psi}(x)\left(i\gamma^{\mu}D_{\mu}-\hat{m}\right)\psi(x)
-\bar{\psi}\left(\sigma+i\gamma_5\vec{\tau}\cdot\vec{\pi}\right)\psi\nonumber\\
&&-\frac{(\sigma^2+\vec{\pi}^2)}{4G_S}+\frac{(V_{\mu}^{a}V^{a\mu}+A_{\mu}^{a}A^{a\mu})}{4G_V}-\frac{B^2}{2}\,,
\end{eqnarray}
where the Euler-Lagrange equation of motion for the auxiliary fields lead to the following constraints:
\begin{eqnarray}\label{NE3b}
\sigma(x)&=&-2G_S \langle \bar{\psi}(x)\psi(x) \rangle ,\nonumber\\
\vec{\pi}(x)&=&-2G_S \langle \bar{\psi}(x)i\gamma_5\vec{\tau}\psi(x) \rangle, \nonumber\\
V_\mu^a(x)&=&-2G_V \langle \bar{\psi}(x)\gamma_\mu\tau^a \psi(x) \rangle,\nonumber\\
A_\mu^a(x)&=&-2G_V \langle \bar{\psi}(x)\gamma_\mu\gamma^5\tau^a \psi(x) \rangle.
\end{eqnarray}

Assuming that the masses of the $u$ and $d$ are the same, $m_u=m_d=m_0$, the constituent quark mass $M$ of these light quarks is given by
\be \label{eq:gapequationmq}
M=m_0-2G_S \langle \bar{\psi}\psi \rangle \,.
\ee

In the NJL model at least in the mean field approximation, the chiral condensate increases with the magnetic field at zero temperature as well as at finite temperature. We will introduce the inverse magnetic catalysis for quark condensate around the critical temperature through three different approaches in Sec. III.

\subsection{The charged $\rho$ meson construction by using the Landau level representation of the  quark propagator}

In the framework of NJL model,  the $\rho$ meson can be expressed as an infinite sum of quark-loop chains  by using the random phase approximation. The $\rho$-meson propagator $D_{ab}^{\mu\nu}(q^2)$ can be obtained from the one  quark loop polarization function $\Pi_{\mu\nu,ab}(q^2)$ via the Schwinger-Dyson equation and takes the form of \cite{He:1997gn}
  \begin{eqnarray}
  \label{DSE}
\left[-iD_{ab}^{\mu\nu}\right]&=&\left[-2iG_V\delta_{ab}g^{\mu\nu}\right] +  \left[-2iG_V\delta_{ac}g^{\mu\lambda}\right]
     \left[-i\Pi_{\lambda\sigma,cd}\right]\left[-iD^{\sigma\nu}_{db}\right],
\end{eqnarray}
where $a, b, c, d$ are isospin indices and $\mu, \nu, \lambda, \sigma$ are Lorentz indices. The one quark loop polarization function $\Pi_{\mu\nu,ab}(x,x')$ is given by
\begin{eqnarray}
\label{polarization}
\Pi^{\mu\nu,ab}(x,x')&=&-i{\rm Tr}[\gamma^\mu\tau^aS_Q(x,x')\gamma^\nu\tau^bS_Q(x',x)].
\end{eqnarray}
Here,  we use the Landau level representation of the quark propagator $S_{Q}(x,x')$\cite{Gusynin:1995nb,Chodos:1990vv}, and it takes the form of
\be\label{Landaulevel}
S_{Q}(x,y)=\exp[\frac{iQe}{2}(x-y)^{\mu}A_{\mu}^{ext}(x+y)]\widetilde{S}(x-y).
\ee
where the Fourier transform of $\widetilde{S}$ is
\be\label{propagator}
\widetilde{S}_Q(k)&=& i\exp(-\frac{\mathbf{k}_{\bot}^2}{|QeB|})\sum_{n=0}^{\infty}(-1)^n\frac{D_n(QeB,k)}{k_0^2-k_3^2-m^2-2|QeB|n},
\ee
with
\be\label{Dn}
D_n(QeB,k)&=&(k^0\gamma^0-k^3\gamma^3+m)\Big[(1-i\gamma^1\gamma^2\mathrm{sign}(QeB))L_n(2\frac{\mathbf{k}_{\bot}^2}{|QeB|})\nonumber\\
&&-(1+i\gamma^1\gamma^2\mathrm{sign}(QeB))L_{n-1}(2\frac{\mathbf{k}^2_{\bot}}{|QeB|})\Big]+4(k^1\gamma^1+k^2\gamma^2)L_{n-1}^1(2\frac{\mathbf{k}^2_{\bot}}{|QeB|}),\nonumber\\
\ee
where $L_n^{\alpha}$ are the generalized Laguerre polynomials and  $L_n=L_n^0$, $L_{n-1}^{\alpha}=0$. Note that $Q$ is the diagonal matrix in the flavor space.

For  charged mesons, the magnetic field violates the Lorentz invariance, so $\Pi^{\mu\nu,ab}(x,x')$ should be written as
\be
   \Pi^{\mu\nu,ab}(x,x')=e^{i\Phi(x_{\bot},x'_{\bot})}\int \frac{d^4q}{(2\pi)^4}e^{-iq(x-x')}\widetilde{\Pi}^{\mu\nu,ab}(\mathbf{q}_{\bot},\mathbf{q}_{||}),
\ee
where $\Phi(x_{\bot},x'_{\bot})$ is the so-called Schwinger phase. If we consider the gauge $A_{\mu}^{ext}=-\delta_{\mu1}Bx_2$ and assume that  $eB\geqslant0$, the Schwinger phase is \cite{Andersen:2014xxa}
\be
   \Phi(x_{\bot},x'_{\bot})=\frac{(x^1-x'^1)(x^2+x'^2)}{2l^2},
\ee
here, $l^2=1/|qeB|$ and $q$ is the electric charge of the meson. In Eq.(\ref{polarization}), the isospin Pauli matrices $\tau^a=\tau^{\pm}, \tau^b=\tau^{\mp}$ represent the charged $\rho^{\pm}$ meson respectively and $\tau^{\pm}=\frac{1}{\sqrt{2}}(\tau^1\pm i\tau^2)$. So Eq.(\ref{polarization}) could be rewritten as
\be\label{Pischwingerphase}
  && \exp \left[i\frac{(x^1-x'^1)(x^2+x'^2)}{2l^2}\right]\int \frac{d^4q}{(2\pi)^4}e^{-iq(x-x')}\widetilde{\Pi}^{\mu\nu}(\mathbf{q}_{\bot},\mathbf{q}_{||})=\nonumber\\
  && -i\exp\left[i\frac{(x^1-x'^1)(x^2+x'^2)}{2l^2}\right]N_ftr_{cs}\Big[\gamma^{\mu}\int\frac{d^4p}{(2\pi)^4}e^{-ip(x-x')}\widetilde{S}(p)\nonumber\\
  &&\times\gamma^{\nu}\int\frac{d^4k}{(2\pi)^4}e^{-ik(x'-x)}\widetilde{S}(k)\Big].
\ee
Considering the momentum conservation i.e. $p=k+q$,  Eq.(\ref{Pischwingerphase}) can be presented in the momentum space
\be
\widetilde{\Pi}^{\mu\nu}(\mathbf{q}_{\bot},\mathbf{q}_{||})=-i\int\frac{d^4k}{(2\pi)^4}N_f{\rm tr}_{cs}\Big[\gamma^{\mu}\widetilde{S}(k+q)\gamma^{\nu}\widetilde{S}(k)\Big].
\ee
In the rest frame of $\rho^{\pm}$ meson, we can obtain the matrix in the Lorentz indices
\be
 \label{eq:matrixrhopm}
\widetilde{\Pi}^{\mu\nu}_{\rho^{\pm}}=\left(\begin{matrix}0&0&0&0\cr0&\widetilde{\Pi}^{11}&\widetilde{\Pi}^{12}&0 \cr0& \widetilde{\Pi}^{21}&\widetilde{\Pi}^{22}&0\cr0&0&0&\widetilde{\Pi}^{33} \end{matrix}\right)
=\left(\begin{matrix}0&0&0&0\cr0&a& ib &0 \cr0& -ib&a&0\cr0&0&0&c \end{matrix}\right),
\ee
where we define $\widetilde{\Pi}^{11}=\widetilde{\Pi}^{22}=a$, $\widetilde{\Pi}^{12}=-\widetilde{\Pi}^{21}=ib$ and $\widetilde{\Pi}^{33}=c$.

In the rest frame of the $\rho^{\pm}$ meson, we can decompose the $\Pi^{\mu\nu}_{ab}$ in the Lorentz and flavor indices as follows:
\begin{eqnarray}
\label{tensordecompose}
\widetilde{\Pi}_{ab}^{\mu\nu}(\mathbf{q}_{\bot},\mathbf{q}_{||})&=& \left [\widetilde{\Pi}_1^2(\mathbf{q}_{\bot},\mathbf{q}_{||}) P_{1}^{\mu\nu}+\widetilde{\Pi}_2^2(\mathbf{q}_{\bot},\mathbf{q}_{||}) P_{2}^{\mu\nu}+\widetilde{\Pi}_3^2(\mathbf{q}_{\bot},\mathbf{q}_{||}) L^{\mu\nu}+\widetilde{\Pi}_4^2(\mathbf{q}_{\bot},\mathbf{q}_{||})u^{\mu}u^{\nu} \right]\delta_{ab},\nonumber\\
\end{eqnarray}
where $u^{\mu}=(1,0,0,0)$ is the 4-velocity in the rest frame of the $\rho$ meson. In Eq.~(\ref{tensordecompose}) the operators $P_{1}^{\mu\nu}$, $P_{2}^{\mu\nu}$ and $L^{\mu\nu}$ are the spin-projection operators on the polarization states of the $\rho$ mesons with projections $s_z= -1$, $s_z= +1$ and $s_z= 0$, respectively:
\bea
P_{1}^{\mu\nu} & = & -\epsilon_2^{\mu}\epsilon_2^{\nu*}, \qquad s_z= -1 \\
P_{2}^{\mu\nu} & = & -\ep_1^{\mu}\ep_1^{\nu*}, \qquad s_z= +1\\
L^{\mu\nu} & = & -b^{\mu}b^{\nu}, \ \qquad s_z= 0\,,
\eea
where $b^{\mu}=(0,0,0,1)$ is a unit vector along the axis of the external magnetic field $\bs B$, and $\ep_1^{\mu}$ and $\ep_2^{\mu} $ are the right-handed and the left-handed polarization vectors, respectively:
\bea
\ep_1^{\mu} = \frac{1}{\sqrt{2}}(0,1, + i,0)\,,
\qquad
\ep_2^{\mu} = \frac{1}{\sqrt{2}}(0,1,-i,0)\,.
\eea
Consequently, the propagator of the $\rho$ meson in Eq.(\ref{DSE}) takes the following form:
 \bea
\widetilde{D}^{\mu\nu}_{ab}(\mathbf{q}_{\bot},\mathbf{q}_{||}) &=&[\widetilde{D}_1(\mathbf{q}_{\bot},\mathbf{q}_{||})P_{1}^{\mu\nu}+\widetilde{D}_2(\mathbf{q}_{\bot},\mathbf{q}_{||})P_{2}^{\mu\nu} +\widetilde{D}_3(\mathbf{q}_{\bot},\mathbf{q}_{||})L^{\mu\nu}+\widetilde{D}_4(\mathbf{q}_{\bot},\mathbf{q}_{||})u^{\mu}u^{\nu}]\delta_{ab},\nonumber\\
\eea
where the propagators of the meson states with different spin polarizations are
\be
\widetilde{D}_i(\mathbf{q}_{\bot},\mathbf{q}_{||})=\frac{2G_V}{1+2G_V {\widetilde{\Pi}}_{i}^2(\mathbf{q}_{\bot},\mathbf{q}_{||})}\,.
\label{eq:Di:q2}
\ee
The poles of the propagators~\eq{eq:Di:q2} give us the gap equations,
\be{\label{Gap}}
1+2G_V {\widetilde{\Pi}}_{i}^2 (\mathbf{q}_{\bot},\mathbf{q}_{||}) = 0,
\ee
that determine the masses of $\rho$ meson with the corresponding spin projections.

Combining the matrix in Eq.(\ref{eq:matrixrhopm}) and the expression in Eq.(\ref{tensordecompose}), the relationship of the matrix
elements and $\Pi_i$ is given by
 \bea
\label{Pi^2}
&&{\widetilde{\Pi}}^2_{1}=-(a+b),  \nonumber\\
&&{\widetilde{\Pi}}^2_{2}=b-a, \nonumber\\
&&{\widetilde{\Pi}}^2_{3}=-c.
\eea
$\widetilde{\Pi}_4^2$ should be zero in the rest frame of $\rho$ meson both in the vacuum and at finite temperature \cite{Das:1994vr},
guaranteed by the Ward identity.

In Ref.\cite{Liu:2014uwa}, the matrix elements $\widetilde{\Pi}^{11}$ and $\widetilde{\Pi}^{12}$ in Eq.\ref{eq:matrixrhopm} are calculated in the vacuum for $\rho^{-}$ meson. If we consider the finite temperature, the matrix elements are given by
\be\label{Pi11}
& & \widetilde{\Pi}^{11}=iN_cN_f\int\frac{d^3k}{(2\pi)^3}\exp\left(-\frac{9\textbf{k}_\bot^2}{2|eB|}\right)
\sum_{k=0}^{\infty}\sum_{p=0}^{\infty}(-1)^{p+k} \nonumber\\&&\left[8\left(L_k(3\frac{\textbf{k}_\bot^2}{|eB|})L_p
(6\frac{\textbf{k}_\bot^2}{|eB|})+L_{k-1}(3\frac{\textbf{k}_\bot^2}
{|eB|})L_{p-1}(6\frac{\textbf{k}_\bot^2}{|eB|})\right)
\right] \nonumber\\
&&\left\{\frac{1}{2}(I_1+I'_1)+\left(\frac{2}{3}|eB|k+\frac{1}{3}|eB|p-\frac{1}{2}M^2_{\rho^-}\right)I_2\right\},
 \ee
 and
 \be\label{Pi12}
 & & \widetilde{\Pi}^{12}= iN_cN_f\int\frac{d^3k}{(2\pi)^3}\exp\left(-\frac{9\textbf{k}_\bot^2}{2|eB|}\right)\sum_{k=0}^{\infty}\sum_{p=0}^{\infty}(-1)^{p+k} \nonumber\\&&\left[8i\left(L_{k-1}(3\frac{\textbf{k}_\bot^2}{|eB|})L_{p-1}
(6\frac{\textbf{k}_\bot^2}{|eB|})-L_k(3\frac{\textbf{k}_\bot^2}{|eB|})L_p
(6\frac{\textbf{k}_\bot^2}{|eB|})\right)\right] \nonumber\\&&\left\{\frac{1}{2}(I_1+I'_1)+\left(\frac{2}{3}|eB|k+\frac{1}{3}|eB|p-\frac{1}{2}M^2_{\rho^-}\right)I_2\right\}.
 \ee
 where
 \be\label{intk0}
 I_1&=&\int\frac{dk_0}{2\pi}\frac{1}{k_0^2-\omega^2_{u,k}},\nonumber\\
 I'_1&=&\int\frac{dk_0}{2\pi}\frac{1}{k_0^2-\omega^2_{d,p}},\nonumber\\
 I_2&=&\int\frac{dk_0}{2\pi}\frac{1}{(k_0^2-\omega_{u,k}^2)((k_0+M_{\rho^-})^2-\omega^2_{d,p})},
  \ee
 with $\omega_{u,k}=\sqrt{\frac{4}{3}|eB|k+k_3^2+M^2}$, $\omega_{d,p}=\sqrt{\frac{2}{3}|eB|k+k_3^2+M^2}$.
 As in Ref.\cite{Rehberg:1995nr}, the integral over $k_0$ can be replaced by Matsubara summation according to the prescription $\int\frac{dk_0}{2\pi}(....)=iT\sum_{m=-\infty}^{\infty}(....)$ and the $I_1, I'_1, I_2$ are given by
 \be\label{matsubara}
 I_1&=&i\left(\frac{n_f(\omega_{u,k}-\mu)+n_f(\omega_{u,k}+\mu)-1}{2\omega_{u,k}}\right),\nonumber\\
 I'_1&=&i\left(\frac{n_f(\omega_{d,p}-\mu)+n_f(\omega_{d,p}+\mu)-1}{2\omega_{d,p}}\right),\nonumber\\
 I_2&=&i\left[\frac{n_f(\omega_{u,k}-\mu)}{2\omega_{u,k}}\frac{1}{(\omega_{u,k}+M_{\rho^+})^2-\omega^2_{d,p}}\right.
 \nonumber\\&&\left.-\frac{n_f(-\omega_{u,k}-\mu)}{2\omega_{u,k}}\frac{1}{(-\omega_{u,k}+M_{\rho^+})^2-\omega^2_{d,p}}\right. \nonumber\\
 &&\left.+\frac{n_f(\omega_{d,p}-
\mu)}{2\omega_{d,p}}\frac{1}{(-M_{\rho^+}+\omega_{d,p})^2-\omega^2_{u,k}}\right.\nonumber\\
&&\left.-\frac{n_f(-\omega_{d,p}-\mu)}{2\omega_{d,p}}\frac{1}{(-M_{\rho^+}-\omega_{d,p})^2-\omega^2_{u,k}}\right],
\ee

\section{The effect of IMC on charged $\rho$ condensation}

In Ref.~\cite{Liu:2015pna}, we have used the MC for chiral condensate to produce the constituent quark mass. In this section, we
consider the effect of the IMC on the charged $\rho$ meson condensate. We introduce three different mechanisms for IMC: Fitting the
lattice data, using the running coupling constant $G_S(eB)$ and introducing the chiral chemical potential $\mu_5(eB)=0.5\sqrt{eB}$.
For numerical calculation, we use the soft cutoff function\cite{Frasca:2011zn}
\bea
&&f_\Lambda=\sqrt{\frac{\Lambda^{10}}{\Lambda^{10}+\mathbf{k}^{2*5}}}, \\
&&f_{\Lambda,eB}^k=\sqrt{\frac{\Lambda^{10}}{\Lambda^{10}+(k_3^2+2|Q eB|k)^5}},
\eea
for zero and nonzero magnetic fields, respectively. We sum up to 30 Landau levels and the results are saturated. As in Ref.\cite{He:1997gn}, we obtain the parameters
 $\Lambda=582$ MeV, $G_S\Lambda^2=2.388$, and $G_V\Lambda^2=1.73$ by reproducing the $\pi$ decay constant $f_{\pi}=95$ MeV, the mass of $\pi$ $m_{\pi}=140$ MeV, $\rho$ mass $M_{\rho}=768$ MeV in the vacuum. Moreover, the vacuum quark mass is $M=458$ MeV and the current quark mass is $m_0=5$ MeV.

\subsection{Fitting the lattice data}

\begin{figure}[!thb]
\centerline{\includegraphics[width=8cm]{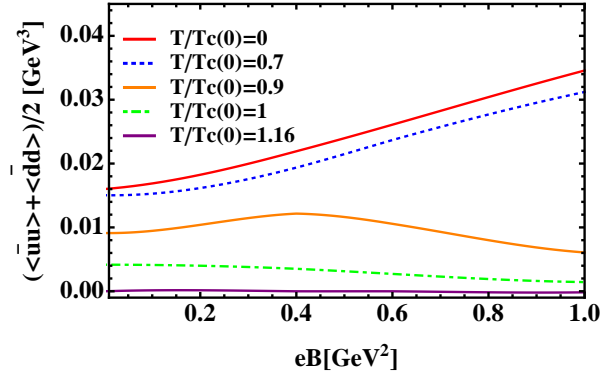}}
\caption{The $eB$ dependence of quark  condensation at fixed T by fitting the lattice data in \cite{Bali:2012zg}.}
\label{fig:condenlattice}
\end{figure}

\begin{figure}[!thb]
\centerline{\includegraphics[width=8cm]{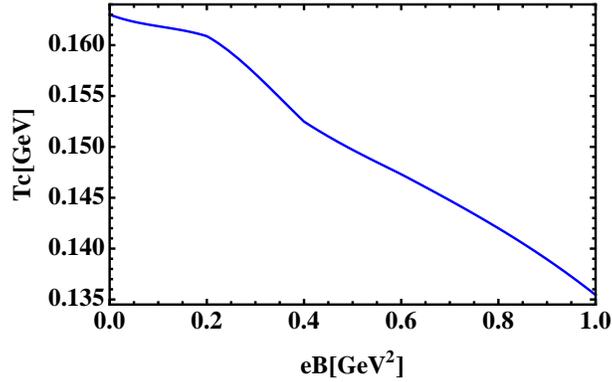}}
\caption{The $eB$ dependence of critical temperature $T_c$ from lattice calculation \cite{Bali:2012zg}.}
\label{fig:Tclatticefit}
\end{figure}

In Ref.\cite{Bali:2012zg,Bali:2011qj}, the lattice group confirmed  the magnetic catalysis at low temperature which is predicted by most of QCD model calculations. Moreover, they firstly observed the inverse magnetic catalysis around the critical temperature $T_c$ of the chiral symmetry restoration, and the dimensionless quantity
 \bea
\label{dimensionless}
\Sigma_{u,d}(B,T)=\frac{2m_{0}}{m^2_{\pi}f^2_{\pi}}[<\bar{\psi}\psi>_{u,d}(B,T)-<\bar{\psi}\psi>_{u,d}(0,0)]+1
\eea
was defined in Ref.\cite{Bali:2012zg}. We combine the lattice data $(\Sigma_u+\Sigma_d)/2$ and $\Sigma_u-\Sigma_d$ with different $T$ and $eB$ in \cite{Bali:2012zg} and
our parameters to reproduce the quark condensation, and the magnetic field dependence of quark condensation is shown in Fig.~\ref{fig:condenlattice}. In Fig.~\ref{fig:Tclatticefit}, we present the $eB$ dependence of critical temperature $T_c$ from lattice calculation. It can be seen that the IMC scenario around $T_c$ has been produced.

\begin{figure}[!thb]
\centerline{\includegraphics[width=8cm]{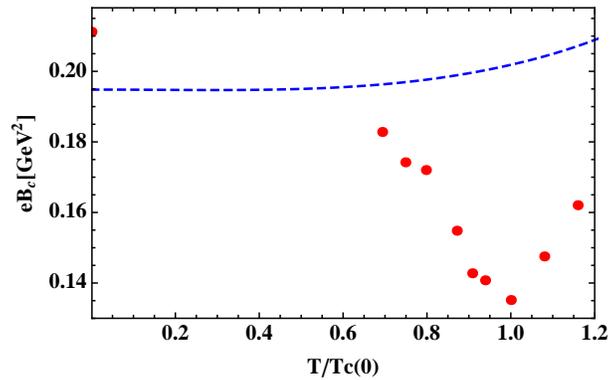}}
\caption{The $T$ dependence of critical magnetic field $eB_c$ by fitting the lattice data in \cite{Bali:2012zg}.
The dashed line is the $eB_c(T)$ by using magnetic catalysis effect for the quark condensation in the NJL model.}
\label{fig:eBcTlattice}
\end{figure}
Using the quark condensation in Fig.~\ref{fig:condenlattice} and solving the gap equation Eq.(\ref{Gap}), we can obtain the critical magnetic field $eB_c$ of polarized charged $\rho^{\pm}$ meson condensation, and the $T$ dependence of $eB_c$ is shown in Fig.~\ref{fig:eBcTlattice}.
At zero temperature, the $eB_c$ of polarized charged $\rho^{\pm}$ meson condensation with IMC effect by fitting lattice data is a little bit larger than its value without IMC, that's because the quark condensate at zero temperature from lattice calculation is a little bit larger than its value in the NJL model \cite{Liu:2015pna}. As shown by the dashed line, if no mechanism for IMC is included, the critical magnetic field $eB_c$ for charged $\rho$ condensation monotonically increases with $T$. However, when the IMC is considered, as shown by dots, the critical magnetic field $eB_c$ decreases with the temperature $T$ firstly, reaches the minimum around $T_c$, then increases with $T$. It can be explicitly seen that the IMC substantially affects the polarized charged $\rho$ condensation
around $T_c$. With IMC effect, the critical magnetic field $eB_c$ for charged $\rho$ condensation drops to $0.13 {\rm GeV}^2$ at $T_c$, and the critical magnetic field $eB_c$ in the temperature region $T<1.3 T_c$ is smaller than its value at $T=0$.

 \subsection{Using the running coupling constant $G_S(eB)$}

\begin{figure}[!thb]
\centerline{\includegraphics[width=8cm]{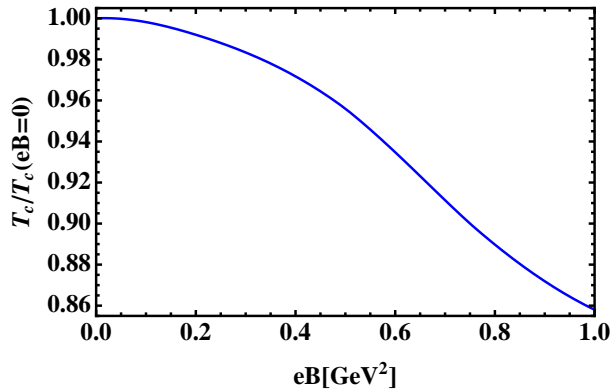}}
\caption{The renormalized critical temperature of the chiral transition as a function of  $eB$ in the NJL model, with the magnetic field dependent coupling $G_s(eB)$ and this corresponds to the Lattice calculation.}
\label{fig:TceBGseB}
\end{figure}

\begin{figure}[!thb]
\centerline{\includegraphics[width=8cm]{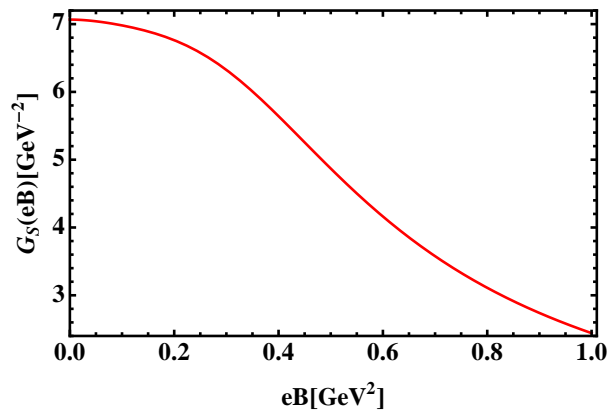}}
\caption{The fitted function $G_S(eB)$ which is used to reproduce the $T_c^{\chi}(eB)$ in the LQCD.}
\label{fig:GseB}
\end{figure}

 In this part, we introduce the running scalar coupling constant $G_S(eB)$ into the NJL model  to calculate the constituent quark mass. Following Ref.\cite{Ferreira:2014kpa}, we fit $G_s(eB)$ in order to reproduce $T_c/T_c(eB=0)$ obtained in Ref.\cite{Bali:2011qj}. As in Ref.\cite{Ferreira:2014kpa}, we present the critical temperature which is used to fit the $G_S(eB)$ in Fig.~\ref{fig:TceBGseB}. We assume that
\bea
\label{GseB}
\frac{G_S(\xi)}{G_S(0)}=\frac{1+a\xi^2+b\xi^3}{1+c\xi^2+d\xi^4}
\eea
where $G_S(0)=G_S$, $\xi=\frac{eB}{\Lambda^2_{QCD}}$ and $\Lambda_{QCD}=300$ MeV. The fitted function $G_S(eB)$ is shown in Fig.~\ref{fig:GseB} and the parameters are $a=0.014056, b=0.00532074, c=0.0281766, d=0.00161148$. It is obviously that the fitted function $G_S(eB)$ should decrease with magnetic field in order to produce the IMC.

\begin{figure}[!thb]
\centerline{\includegraphics[width=8cm]{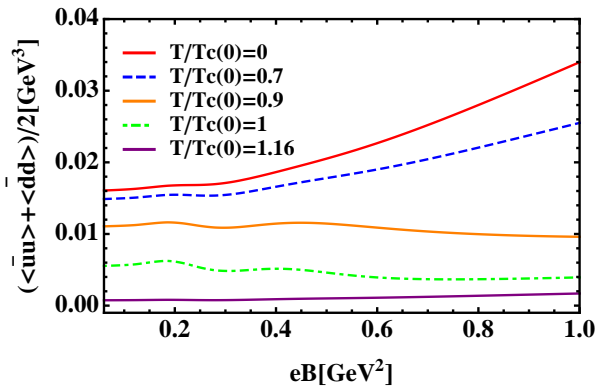}}
\caption{The $eB$ dependence of quark  condensation by using the running coupling constant $G_S(eB)$.}
\label{fig:Gs(eB)conden}
\end{figure}

\begin{figure}[!thb]
\centerline{\includegraphics[width=8cm]{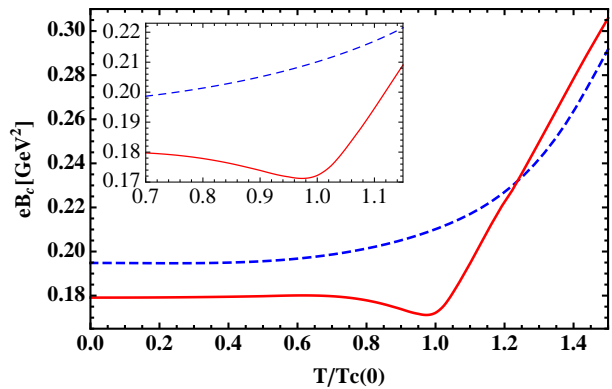}}
\caption{The temperature $T$ dependence of critical magnetic field $eB_c$ by using the running coupling constant $G_S(eB)$ to reproduce the IMC. The dashed line is the case of MC for quark condensation.}
\label{fig:eBcTGs(eB)}
\end{figure}

Then we  solve the dynamical quark mass in the NJL model with the running coupling constant $G_S(eB)$. The one-loop level effective potential of the model is given by
 \begin{eqnarray} \label{eq:effectivepotential}
 \Omega&=&\frac{(m-m_0)^2}{4G_S(eB)}+\frac{B^2}{2}-3\sum_{q_f\in\{\frac{2}{3},-\frac{1}{3}\}}\frac{|q_feB|}{\beta}\sum_{p=0}^{+\infty}\alpha_p\int_{-\infty}^{+\infty}\frac{dp_3}{4\pi^2} \nonumber\\ &&\left\{\beta E_q+\ln \left(1+e^{-\beta(E_q+\mu)}\right)+\ln \left(1+e^{-\beta(E_q-\mu)}\right)\right\},\nonumber\\
 \end{eqnarray}
where $\beta=1/T$ and $\alpha_p=2-\delta_{p0}$ is the spin degeneracy factor. To minimize the effective potential, we can determine the dynamical quark mass. By using Eq.(\ref{NE3b}), we obtain the quark condensation as shown in Fig.~\ref{fig:Gs(eB)conden} and the critical magnetic field $eB_c(T)$  in Fig.~\ref{fig:eBcTGs(eB)}.

As we can see, the critical magnetic field as a function of the temperature $eB_c(T)$ shows similar behavior as in the last subsection. We can see that
the IMC substantially affects the polarized charged $\rho$ condensation around $T_c$. If no mechanism for IMC is included, the critical magnetic field $eB_c$ for charged $\rho$ condensation monotonically increases with $T$, which can be read from the dashed line. However, when the IMC is considered, the critical magnetic field $eB_c$ decreases with the temperature $T$ firstly, reaches the minimum around $T_c$, then increases with $T$.
With IMC effect, the critical magnetic field $eB_c$ for charged $\rho$ condensation drops to $0.17 {\rm GeV}^2$ around $T_c$, and the critical magnetic field $eB_c$ in the temperature region $T<1.3 T_c$ is smaller than its value when no IMC is considered. It is also noticed that the critical magnetic field $eB_c$ with IMC below $1.2T_c$ is lower than its corresponding value in the original NJL model \cite{Liu:2015pna} when no IMC
is considered, this is because in the case of IMC, the coupling constant decreases with the magnetic field as can be seen from Fig. \ref{fig:GseB}.

 \subsection{Introducing the chiral chemical potential $\mu_5(eB)=0.5\sqrt{eB}$}

\begin{figure}[!thb]
\centerline{\includegraphics[width=8cm]{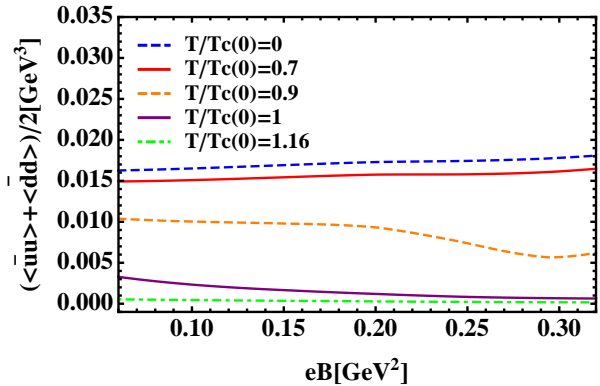}}
\caption{The $eB$ dependence of quark  condensation by introducing the $\mu_5(eB)=0.5\sqrt{eB}$.}
\label{fig:mu5conden}
\end{figure}

\begin{figure}[!thb]
\centerline{\includegraphics[width=8cm]{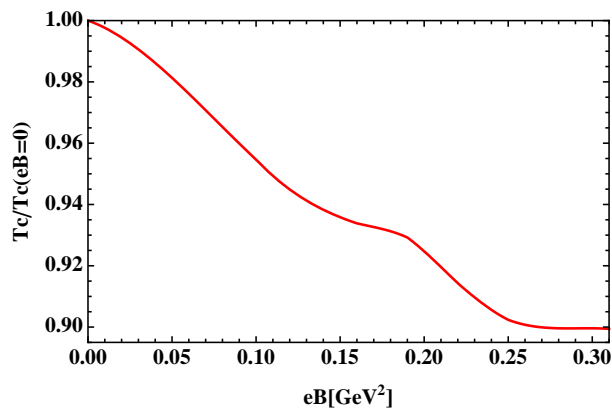}}
\caption{The critical temperature of the chiral transition as a function of $eB$ by introducing the $\mu_5(eB)=0.5\sqrt{eB}$ in the NJL model.}
\label{fig:mu5eBTc}
\end{figure}

\begin{figure}[!thb]
\centerline{\includegraphics[width=8cm]{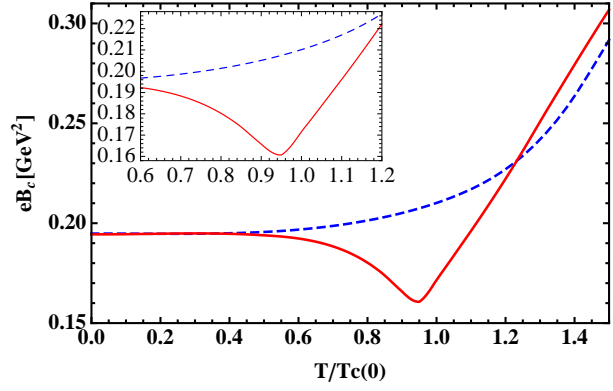}}
\caption{The temperature $T$ dependence of the critical magnetic field $eB_c$ by introducing $\mu_5(eB)=0.5\sqrt{eB}$ to reproduce the IMC.
The dashed line is the case of MC for quark condensation.}
\label{fig:eBcTmu5}
\end{figure}

 In Refs.\cite{Yu:2014sla,Chao:2013qpa}, the chiral chemical potential $\mu_5(eB)$  is introduced to reproduce the IMC phenomenon. In the following, we will use $\mu_5(eB)=0.5\sqrt{eB}$ to produce the IMC and corresponding quark mass. The $eB$ dependence of  the quark condensation is shown in Fig.~\ref{fig:mu5conden} and the critical temperature as a function of $eB$ is shown in Fig.~\ref{fig:mu5eBTc}.
It is evident that the IMC around $T_c$ is reproduced. Then, we solve the critical magnetic field $eB_c$ for charged $\rho$ condensation by using the gap equation Eq.(\ref{Gap}) and the function $eB_c(T)$ is shown in Fig.~\ref{fig:eBcTmu5}.

As we can see, the critical magnetic field as a function of the temperature $eB_c(T)$ shows similar behavior as in the last two subsections, and
the IMC substantially affects the polarized charged $\rho$ condensation around $T_c$. If no mechanism for IMC is included, the critical magnetic field $eB_c$ for charged $\rho$ condensation monotonically increases with $T$, which can be read from the dashed line. However, when the IMC is considered, the critical magnetic field $eB_c$ decreases with the temperature $T$ firstly, reaches the minimum $0.16 {\rm GeV}^2$ around $T_c$,
then increases with $T$, and the critical magnetic field $eB_c$ in the temperature region $T<1.3 T_c$ is smaller than its value when no IMC is considered.

\section{The decay of the strong magnetic field}

In the last section, we investigate the temperature  $T$ dependence of $eB_c$ by considering the Inverse Magnetic Catalysis for chiral condensate. By using three different mechanisms for IMC, we find that the critical magnetic field $eB_c(T)$ shows the same behavior around $T_c$, i.e.  $eB_c$ decreases with the temperature and drops to its minimum around $T_c$ and then increases with the temperature $T$. It is also noticed that in the temperature region $T<1.3 T_c$, the critical magnetic field $eB_c$  is smaller than its value when no IMC is considered. Our calculation shows that the charged $\rho$ condensation can exist in the temperature region of $1-1.5 T_c$ with rather small critical magnetic field $eB_c\sim 0.15-0.3 {\rm GeV}^2$. It is suggested that high temperature superconductor might be created in non-central heavy ion collisions at LHC energies. If high temperature superconductor is formed in the early stage of non-central heavy ion collisions at LHC energies, it should have rather large electric conductivity. It has been shown in Refs.~\cite{Gursoy:2014aka,Li:2016tel} that a finite electric conductivity substantially delays the decay of strong magnetic field. Therefore we investigate how high temperature superconductor will affect the decay of strong magnetic field created in non-central heavy ion collisions.

Following Ref.~\cite{Gursoy:2014aka}, we estimate the magnetic field in the center-of-mass frame. By solving the wave function, the $y$-component
of magnetic field $\vec B$ is given by
  \bea\label{B}
   eB^+_y(\tau,\eta,x_\bot,\phi) &=& \alpha\sinh(Y_b)(x_\bot \cos\phi -  x'_\bot \cos\phi') \nonumber
   \\&&\frac{(\frac{\sigma_E|\sinh(Y_b)|}{2} \sqrt{\Delta} +1)}{\Delta^{\frac{3}{2}}} e^A,
  \eea
where $\sigma_E$ is the electric conductivity, $\alpha=e^2/(4\pi)$ is the electromagnetic coupling, $Y_b=$arctanh$(\beta)$ is the rapidity,
and we have defined
 \bea\label{ADelta}
 A&=&\frac{\sigma_E}{2}(\tau\sinh(Y_b)\sinh(Y_b-\eta)-|\sinh(Y_b)\sqrt{\Delta}|)\nonumber\\
 \Delta&=&\tau^2\sinh^2(Y_b-\eta)+x^2_{\bot}+x'^2_{\bot}-2x_{\bot}x'_{\bot}\cos(\phi-\phi')
\eea

Because high temperature superconductor, if it is can be formed, is formed after the non-central collision, so we assume that the electrical
conductivity of the QGP $\sigma_E$ is time-dependent and takes the following form:
\begin{equation}\label{sigma}
\sigma_E(\tau)=\left\{
\begin{aligned}
&k\tau   \ \ \ \ \ \ \ \ \ \ \ \   \sigma_E<\sigma_E^{max}\\
&\sigma_E^{max}   \ \ \ \ \ \ \ \ \    \sigma_E\geqslant\sigma_E^{max}\\
\end{aligned}
\right.
\end{equation}
Here $k$ is the growing rate and $\sigma_E^{max}/k$ gives the formation time for the high temperature superconductor.

\begin{figure}[!thb]
\centerline{\includegraphics[width=11cm]{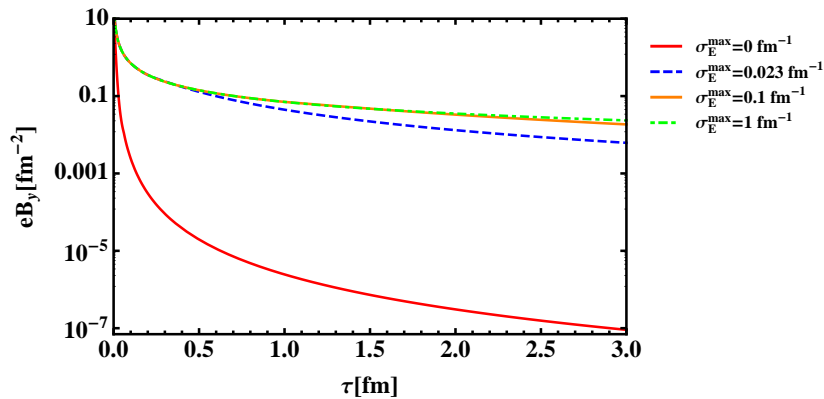}}
\caption{The time-dependent magnetic field $eB_y$ with different $\sigma_E^{max}$.}
\label{fig:eBdecaysigmachange}
\end{figure}

With simple assumption, the protons distribution in either the + moving or the - moving nucleus onto the transverse plane is given by
\bea\label{distribution}
n_{\pm}(x_\bot)=\frac{3}{2\pi R^3}\sqrt{R^2}-(x^2_{\bot}\pm bx_{\bot}\cos(\pi)+\frac{b^2}{4})
\eea
In a collision, while the impact parameter is  $b\neq 0$ and the $+$ and $-$ moving spectators
are each located in a crescent-shaped region of the
$\vec x'_\perp$-plane,
one can write the total magnetic field produced by all the spectators as~\cite{Kharzeev:2007jp}
\bea\label{totBys}
eB_{y,s} &=&  -Z\int_{-\frac{\pi}{2}}^{\frac{\pi}{2}} d\phi' \int_{x_{\rm in}(\phi')}^{x_{\rm out}(\phi')} dx'_{\bot} x'_{\bot} n_-(x'_{\bot}) \\
&&\hspace{-0.2in}\times\left(eB^+_{y}(\tau,\eta,x_{\bot},\pi-\phi) + eB^+_{y}(\tau,-\eta,x_{\bot},\phi)\right)\, ,
\eea
with $B_y^+$defined in Eq.(\ref{B}).
Here $x_{\rm in}$ and $x_{\rm out}$ are the endpoints of the $x'_{\bot}$ integration
regions that define the crescent-shaped loci where one finds either $+$ movers
or $-$ movers but not both.  They are given by
\bea\label{xpm}
x_{\rm in/out}(\phi') = \mp \frac{b}{2}\cos(\phi') + \sqrt{R^2-\frac{b^2}{4}\sin^2(\phi')}\, .
\eea
At the LHC, we assume that  $Y_b=Y_0=8$, $R=7$ fm, $b=7$ fm, $\phi=0$, $x_{\bot}=0$, $\eta=0$ and $Z=82$ and the details are in Ref.~\cite{Gursoy:2014aka}. The strong magnetic field with magnitude $eB_y=10  {\rm fm}^{-2}=0.4 {\rm GeV}^2$ is created through non-central heavy ion collision. We present numerical results of the decay of $eB_y(\tau)$ with different $\sigma_E^{max}$ in Fig.~\ref{fig:eBdecaysigmachange} by choosing  $k=0.1$ fm$^{-2}$.

With the simple assumption of  the time-dependent electric conductivity in Eq.(\ref{sigma}),  it is found that the conducting medium with
$\sigma_E$ effectively delays the decay of the magnetic field. At later stage, the larger the $\sigma_E$ is, the slower the decay of the magnetic field.
It is also worthy of mentioning that if one chooses a constant $\sigma_E$ starting from $\tau=0$ as in \cite{Gursoy:2014aka},
even a small electric conductivity will make the magnetic field at $\tau=0$ drops to a small value. Our results suggests that a growing electric
conductivity starting from zero at $\tau=0$ will substantially delay the decay of the magnetic field, which is helpful for the formation of the high
temperature superconductor.

\section{Conclusion}

In this paper, we have investigated the charged $\rho$ condensation in an external magnetic field at finite temperature in the NJL model. When the $\rho$ mass decreases with magnetic field and becomes zero, it indicates there exits charged $\rho$ condensation. The magnetic field when the charged $\rho$ becomes massless is named as the critical magnetic field $eB_c$. In the NJL model, the mesons are constructed by a infinite sum of quark-loop chains by using the random phase approximation and we calculate to the leading order of $1/N_c$ expansion for $\rho$ meson.

We have studied the temperature $T$ dependence of $eB_c$ by considering the IMC for quark chiral condensate. We use three different ways to reproduce the IMC, i.e. fitting the Lattice data, using the running  scalar coupling constant $G_S(eB)$ and considering the chirality imbalance $\mu_5(eB)=0.5\sqrt{eB}$, respectively. It is found that, in all three cases, the IMC substantially affects the polarized charged
$\rho$ condensation around the critical temperature $T_c$ of chiral phase transition. If there is no IMC effect included, the critical magnetic field  $eB_c$ for charged $\rho$ condensation increases monotonically with the temperature. However, when the effect of IMC is considered, the critical magnetic field $eB_c$ for charged $\rho$ condensation decreases with the temperature firstly and reaches to a minimum value around the critical
temperature $T_c$ and then increases with the temperature. The critical magnetic field $eB_c$ with IMC in the temperature region $T<1.2 T_c$ is smaller than its value if no IMC is considered. Our calculation shows that the charged $\rho$ condensation can exist in the temperature region of $1-1.5 T_c$ with critical magnetic field $eB_c\sim 0.15-0.3 {\rm GeV}^2$, which can be regarded as high temperature superconductor, and such a high temperature superconductor might be formed in non-central heavy ion collisions at LHC energies. We also show that a growing electric conductivity in early stage of non-central heavy-ion collisions substantially delays the decay of strong magnetic field, which is helpful for the formation of the high temperature superconductor.

It is worth to mention that we have done the research in the background without $\rho$ condensation. It is interesting to investigate the conductivity of charged $\rho$ condensation by introducing the charged polarized $\rho^{\pm}$ condensate in the Lagrangian in the future.

\section{Acknowledge}
We thank J.F. Liao for valuable discussions. This work is supported by the NSFC under
Grant No. 11275213, and 11261130311(CRC 110 by DFG and NSFC), CAS key project KJCX2-EW-N01, and Youth
Innovation Promotion Association of CAS. L. Yu is partially supported by China Postdoctoral Science Foundation under Grant
No.2014M550841 and the Seeds Funding of Jilin University.

\end{document}